\documentclass[amssymb,amsmath,
longbibliography,
aps,
reprint,
superscriptaddress
]{revtex4-1}%aip
\usepackage{graphicx}
\usepackage{textcomp}
\usepackage[update]{epstopdf}
\usepackage{xfrac} %slanted fraction \sfrac{a}{b}
\usepackage[justification=RaggedRight, textfont={sf, small}]{caption} %edit figure caption formating
\begin{document}

\title{Rapid high-fidelity spin state readout in Si/SiGe quantum dots via radio-frequency reflectometry}

\author{Elliot J. \surname{Connors}}
\thanks{These authors contributed equally.}

\author{JJ \surname{Nelson}}
\thanks{These authors contributed equally.}

\affiliation{Department of Physics and Astronomy, University of Rochester, Rochester, NY 14627}

\author{John M. \surname{Nichol}}
\email{john.nichol@rochester.edu}

\affiliation{Department of Physics and Astronomy, University of Rochester, Rochester, NY 14627}

\date{\today}

\begin{abstract}
Silicon spin qubits show great promise as a scalable qubit platform for fault-tolerant quantum computing. 
However, fast high-fidelity readout of charge and spin states, which is required for quantum error correction, has remained elusive. 
Radio-frequency reflectometry enables rapid high-fidelity readout of GaAs spin qubits, but the large capacitances between accumulation gates and the underlying two-dimensional electron gas in accumulation-mode Si quantum dot devices, as well as the relatively low two-dimensional electron gas mobilities, have made radio-frequency reflectometry challenging in these platforms. 
In this work, we implement radio-frequency reflectometry in a Si/SiGe quantum dot device with overlapping gates by making minor device-level changes that eliminate these challenges.  We demonstrate charge state readout with a fidelity above $99.9\%$ in an integration time of $300~\text{ns}$. We measure the singlet and triplet states of a double quantum dot via both conventional Pauli spin blockade and a charge latching mechanism, and we achieve maximum fidelities of $82.9\%$ and $99.0\%$ in $2.08~\mu$s and $1.6~\mu$s integration times, respectively. We also use radio-frequency reflectometry to perform single-shot readout of single-spin states via spin-selective tunneling in microsecond-scale integration times. 
\end{abstract}


\maketitle

\section{Introduction}

Electron spins in gated Si quantum dots are promising qubits because they possess long coherence times, which enable high-fidelity gate operations~\cite{veldhorst2014addressable, tyryshkin2012electron, morello2010single, yoneda2018quantum, eng2015isotopically, maune2012coherent, watson2018programmable, veldhorst2015two,xue2019benchmarking}.
In the future, quantum error correction will require a large number of physical qubits and the ability to measure and correct qubits quickly~\cite{vandersypen2019quantum,zhang2018semiconductor, fowler2012surface}. 
The fabrication of Si spin qubits leverages existing commercial technologies, and the production of large numbers of qubits seems within reach. Moreover, current architectures are compatible with one- and two-dimensional qubit arrays~\cite{zajac2016scalable,Mortemousque2018,Mukhopadhyay2018}.
However, implementing readout methods that are simultaneously fast, high-fidelity, and scalable has been challenging in these systems.

Readout of electron spins in quantum dots is usually performed via spin-to-charge conversion together with an external charge sensor~\cite{simmons2011tunable,thalakulam2011single,field1993,elzerman2003few,pla2012single,zajac2016scalable,broome2017high,maune2012coherent} or gate-based dispersive sensing techniques~\cite{colless2013dispersive,pakkiam2018single,zheng2019rapid, urdampilleta2019gate,west2019gate}.
Gate-based dispersive sensing does not require an additional charge sensor and is therefore inherently scalable, but it is often less sensitive than charge sensing. Charge sensing is easy to implement, sensitive, and compatible with linear qubit arrays, which have emerged as key elements of near-term spin-based quantum information processors~\cite{zajac2016scalable,volk2019qubyte,kandel2019,qiao2019}.

External charge sensors, such as quantum point contacts or quantum dots, can be used for both baseband or radio-frequency (rf) readout.
In the former case, high-bandwidth baseband readout can be achieved, but it requires low-noise cryogenic preamplifiers and careful wiring to minimize stray capacitance~\cite{vink2007,jones2019spin}. In the latter case, rf reflectometry achieves high-bandwidth readout by incorporating the charge sensor into an impedance matching tank circuit~\cite{schoelkopf1998radio,reilly2007fast,taskinen2008}.
Changes to the electrostatic potential of the charge sensor alter its conductance and therefore generate measurable changes to the reflection coefficient of the circuit. This technique enables fast and high-fidelity readout in GaAs quantum dots~\cite{reilly2007fast, barthel2009rapid,barthel2010fast,higginbotham2014coherent}. It is also easy to implement and enables frequency multiplexing for multi-qubit readout.

Radio-frequency reflectometry has successfully been applied to Si donor-based devices~\cite{villis2014,hile2015radio}. 
However, accumulation-mode Si devices present two main challenges to rf reflectometry.
First, accumulation-mode devices can incur sizeable capacitances of order $10^{-12}-10^{-11}$ F if large-area accumulation gates are used. 
This is much larger than typical capacitances in reflectometry circuits, and it can negatively impact the performance of the tank circuit~\cite{taskinen2008}. 
Second, Si devices often have lower mobilities than GaAs devices. 
These lower mobilities generate excess resistance in the two-dimensional electron gas (2DEG), diminishing the sensitivity of the charge sensor. 
Successful rf reflectometry circuits must be optimized such that both the incurred capacitance and excess resistance are sufficiently small.
For example, one cannot simply reduce the size of the accumulation gate arbitrarily, because that would increase the excess resistance. 
Previous work on rf reflectometry in silicon, including recent clever circuit modification strategies to circumvent the large capacitance~\cite{angus2008silicon,yoneda2018quantum,ares2016,wang2013charge, volk2019fast}, have shown promising results.
However, rapid high-fidelity spin state readout via rf reflectometry in silicon remains challenging.

Here, we implement high-fidelity charge- and spin-state sensing via rf reflectometry in a Si/SiGe quantum-dot device with overlapping gates. We eliminate the problems discussed above by making minor device-level changes. These changes are easy to implement and preserve the scalability of the overlapping-gate architecture. We demonstrate high-fidelity charge and singlet-triplet readout in submicrosecond integration times, and we use rf reflectometry to implement microsecond-scale single-spin readout.

\begin{figure}[t!]
\includegraphics[width=8.6cm]{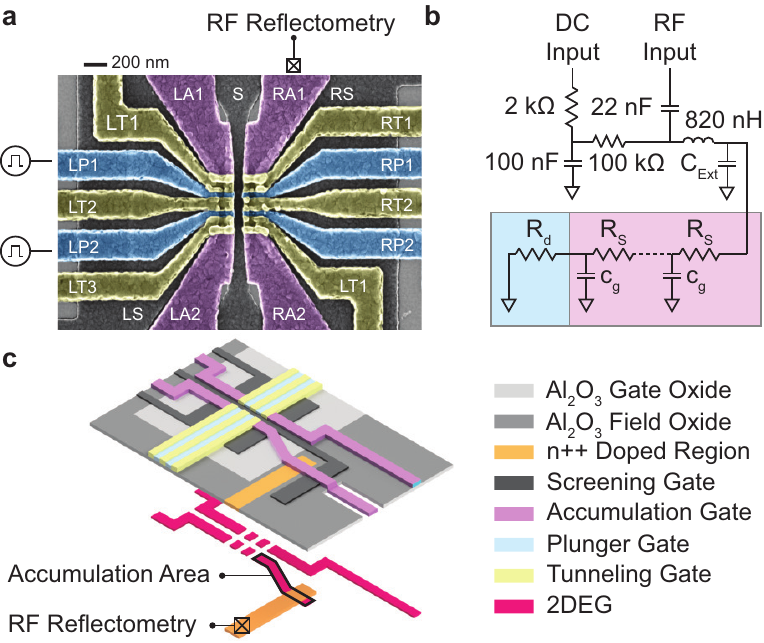}
\caption*{
\textbf{Fig. 1} Device and design.
\textbf{a}~False color scanning electron microscope image of a device with a nominally identical geometry to the ones tested. 
Gates are colored according to their purpose, with accumulation, screening, plunger, and tunneling gates shown in purple, black, blue, and yellow, respectively. 
The rf reflectometry circuit is connected to an ohmic contact to the 2DEG indicated by a square with an "x" in it.
\textbf{b}~Schematic of the rf reflectometry circuit.
The on-chip portion of the circuit is highlighted where the region in pink is the contribution from the 2DEG under the accumulation gate, and represents a distributed network of $R_{S}$ and $c_{g}$ components where $C_G=\sum c_g$ is the total gate capacitance.
The region in blue is the contribution from the dot, and $R_d$ is the resistance of the dot.
\textbf{c}~Schematic of the device used showing the underlying 2DEG formed in typical device operation.
The Accumulation Area, the region of the 2DEG which corresponds to the reflectometry circuit, is outlined in black.
}\label{fig:device}
\end{figure}

\section{Device Design}

The quantum dot devices in this report are fabricated on an undoped Si/SiGe heterostructure with an 8-nm-thick Si quantum well approximately 50~nm below the surface. 
Voltages applied to three overlapping layers of aluminum gates are used to confine electrons in up to four quantum dots~\cite{angus2007gate,zajac2015reconfigurable} (Fig 1a). 

We optimized a quantum dot device for rf reflectometry through the following design rules. 
First, the capacitance between the accumulation gate associated with the readout circuit and the 2DEG, $C_G$, should satisfy $C_G < 5 \times 10^{-14}$ F. 
Second, the total resistance of the path between the doped region associated with the readout circuit and the sensor dot, $R_T=\sum R_S$, should satisfy $R_T < 15 R_S$, where $R_S$ is the resistance per square of the 2DEG. 
Satisfying this condition likely ensures that $R_T \ll R_d$ for typical accumulation gate voltages and densities in Si/SiGe. 
Here $R_d$ is the resistance of the sensor dot.
We arrived at these empirical design rules by evaluating several prototype devices (Table I). 
As described below, the final device design has vastly reduced values of $C_G$ and $R_T$ compared to initial devices.
%These design rules were come to empiracally through the evaluation of various prototype device designs, and are understood to reduce $C_G$ and $R_T$ to values suitable for rf reflectometry.

To accommodate these design rules, the $n^{++}$ region in the reflectometry circuit extends to within 10 $\mu$m of the sensor dot, which helps to reduce $R_T$ (Fig. 1c). We use a screening gate~\cite{west2019gate}, which runs underneath the accumulation gate associated with the reflectometry circuit, RA1, to reduce $C_G$ as much as possible.
Additionally, we remove the quantum well everywhere under RA1 except between the $n^{++}$ region and the dot via a dry etch process to ensure a minimization of $C_G$.
The accumulation gate corresponding to the rf channel has a $12\text{-\ensuremath{\mu}m}^{2}$ area consisting of roughly $8$ squares between the screening gate and the dot.
The device has a 15-nm-thick Al\textsubscript{2}O\textsubscript{3} gate-oxide layer and a 30-nm-thick Al\textsubscript{2}O\textsubscript{3} field-oxide layer. 
While a thinner gate-oxide layer reduces charge noise~\cite{connors2019low}, it increases $C_G$. 
Although a matching capacitor can improve sensitivity of reflectometry in devices with large $C_G$~\cite{ares2016}, the large $R_T$, which is distributed with $C_G$, seemed to prevent success with this approach in our devices. 
The device optimization described above represents a relatively simple method to implement rf reflectometry, and it should be widely applicable to most accumulation-mode Si quantum dot devices.  

\begin{table}[b!]
\begin{ruledtabular}
\begin{tabular}{ccccc}
\scriptsize{Device} & \scriptsize{Accumulation Area} & \scriptsize{$N_S$} & \scriptsize{Sensitivity} & \scriptsize{$F_C$} \\
& \scriptsize{$\left(\mu\text{m}^2\right)$} & & \scriptsize{$\left(\frac{dDAQ}{dV_{RP1}}\right)$} & \scriptsize{$\left(T_{int}=1~\mu\text{s}\right)$} \\
\hline
\hline
FET & $\sim130$ & 9 & - & - \\
QD 0 & $\sim5000$ & $\sim30$ & 0 & - \\
QD 1 & 83 & 11 & 5 & - \\
QD 2 & 30 & 8 & 25 & $94.4\%$ \\
QD 3 & 12 & 8 & 27 & $>99.9\%$ \\
\end{tabular}
\end{ruledtabular}
\caption*{
\textbf{Table I} 
Parameters and performance metrics of devices tested.
Accumulation Area is the area of the portion of the accumulation gate corresponding to the reflectometry circuit that accumulates a 2DEG beneath it and thus incurs a gate-to-2DEG capacitance, as shown in Figure 1c.
$N_S$ is the number of squares in the accumulation area.
Sensitivity is the slope of the conductance peak used for charge sensing.
We report here a typical achievable value of the sensitivity, and not the maximum value measured.
$F_C(T_{int}=1~\mu\text{s})$ is the charge state readout fidelity as defined in Equation 3 with an integration time $T_{int}=1~\mu\text{s}$.
QD 0 is a four quantum dot device without any design optimization for rf reflectometry.
QD 1, 2 and 3 are four quantum dot devices at various iterations of rf reflectometry optimization.
Measurements reported in sections III-V of this work were made on QD 3.
}
\end{table}

\begin{figure*}[t!]
\includegraphics[width=17.2cm]{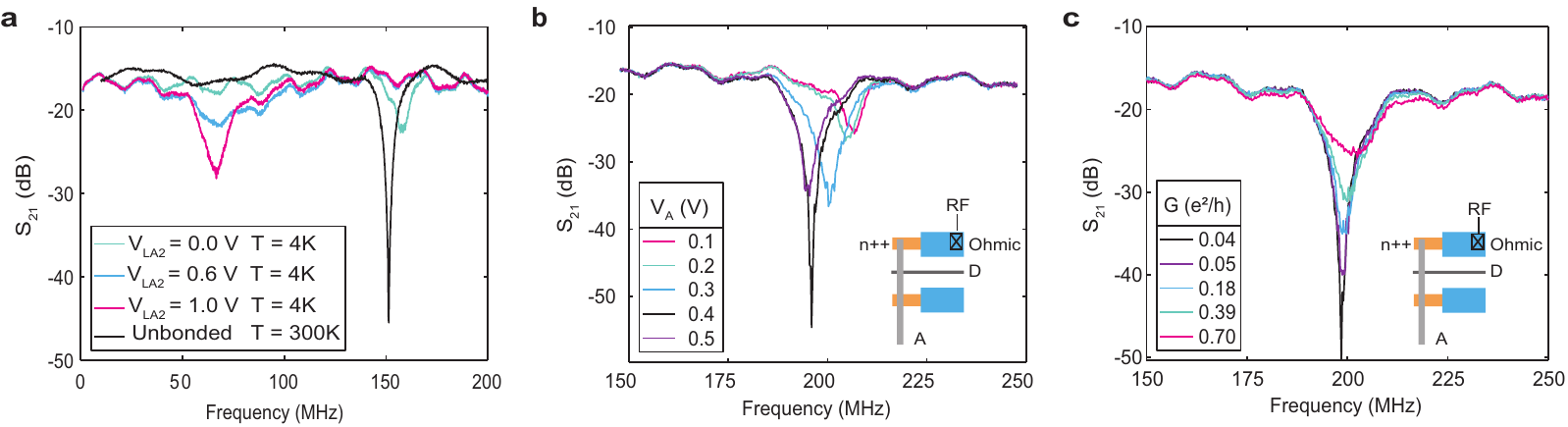}
\caption*{
\textbf{Fig. 2} Rf reflectometry performance in prototype devices.
\textbf{a} Measurement of $S_{21}$ in a quantum dot device that is not optimized for rf reflectometry (QD 0) made using a network analyzer connected to the ohmic contact corresponding to LA2 via a bias tee.
The reflected rf signal is sensitive to the accumulation gate voltage, but no change is observed when the tunneling gates are used to modulate the channel conductivity.
The large shift in the resonance frequency from $\sim$150~MHz to $\sim$70~MHz occurs when the accumulation gate is held above threshold thereby inducing a large $C_G$.
\textbf{b} Demonstration of in situ impedance matching in a two-gate FET.
$V_{A}$ is the voltage on the accumulation gate (A) with the depletion gate (D) held below threshold.
\textbf{c} Response of the reflected rf signal as the channel conductance is varied using the depletion gate.
The -16~dB offset in \textbf{a}-\textbf{c} arises from the combination of room temperature attenuation and amplification used on the send and return signals, respectively.
The oscillations in the baseline arise from stray reflections in the measurement setup. 
The insets in \textbf{b} and \textbf{c} show a schematic of the two-gate FET tested.
All measurements in \textbf{b} and \textbf{c} were made at approximately 4~K.
}\label{fig:fet}
\end{figure*}

The above mentioned design criteria were evaluated in a prototype two-gate field effect transistor (FET) consisting of an accumulation gate spanning source and drain contacts and a depletion gate used to modulate the channel conductance.
The relevant geometries of the FET device were designed to mimic the rf reflectometry circuit in a quantum dot device.
To evaluate the rf reflectometry performance, we cooled the prototype device to approximately 4~K and measured $S_{21}$ using a network analyzer connected to the device via a bias tee and directional coupler.
For comparison, Figure 2a shows the rf response of a quantum dot device not optimized for rf reflectometry while 2b and 2c are the results of measurements performed on the FET. 
Figure 2b shows the response of the reflected signal to the accumulation gate voltage while using the depletion gate to fully suppress the channel conductance. 
The observed resonance is strongly affected by the gate voltage, demonstrating the capability of using an accumulation gate for in-situ impedance matching.
Once the circuit is matched via adjustments to the accumulation gate voltage, the reflected rf signal is sufficiently sensitive to changes in the channel conductance induced by the depletion gate (Fig. 2c). 
A summary of relevant design parameters and performance metrics for select devices used to prototype the rf reflectometry optimization, including revisions of quantum dot devices, is given in Table I.
The design rules and all results in the following sections pertain to device QD 3.

\section{Charge State Readout}

We cool our rf reflectometry optimized device (QD 3 in Table I) in a dilution refrigerator to a base temperature of approximately 50~mK and tune the gate voltages to form a sensor quantum dot under plunger gate RP1.
We apply an rf excitation at $224$~MHz to the ohmic contact corresponding to RA1, which is part of the impedance-matching circuit.
The rf carrier is generated at 10~dBm and sees 36~dB, 20~dB, and 13~dB of attenuation at room temperature, 1.5~K, and 50~mK, respectively.
The circuit also consists of an 820-nH surface mount inductor and the stray capacitance of the device, $C_{Ext}$ (Fig. 1b).
The reflected rf signal is amplified by 38~dB via a Cosmic Microwave Technologies CITLF3 cryogenic amplifier at approximately 4~K and an additional 54~dB via a Narda-MITEQ AU-1565 amplifier at room temperture before it is digitized on an AlazarTech ATS9440 data acquisition card (DAQ)~\cite{reilly2007fast}.
During the tune-up process, we adjust the accumulation gate voltage, which affects both $C_G$ and $R_T$, to optimize the sensitivity of the circuit~\cite{angus2008silicon}.
We observe a strong modulation in the reflected signal as the plunger gate sweeps across conductance peaks (Fig. 3a). %The Supplemental Material~\cite{rfrosupp} contains further details about the reflectometry circuit.

To perform charge sensing, we set the plunger gate voltage to the side of a conductance peak such that the reflected rf signal is sensitive to small changes in the electrochemical potential of the dot.
We then tune the gates on the left side of the device to form a double quantum dot, and acquire charge stability diagrams by measuring the reflected rf signal while varying the voltages on plunger gates LP1 and LP2 (Fig. 3b).

We quantify the charge state readout performance by tuning the left-side double dot to the (1,0) occupancy, where the $(i,j)$ notation refers to the charge configuration with $i(j)$ electrons in the dot under gate LP1(LP2). We adjust the tunnel barrier coupling the dot under LP1 to its reservoir such that the tunneling rate is of order 10~Hz.
We set the voltage of the plunger gate directly on the (0,0)-(1,0) transition and acquire a time series of the reflected signal, and we resolve individual charge tunneling events~\cite{zajac2016scalable} (Fig. 3c).
We fit a histogram of the data (Fig. 3d) to a function $G(V)=g(V|A_0,\mu_0,\sigma_0)+g(V|A_1,\mu_1,\sigma_1)$ where 
\begin{equation}\label{eq:gauss}
g(V|A_i,\mu_i,\sigma_i)=\frac{A_i}{\sqrt{2\pi\sigma_i^2}}e^{-\frac{\left( V-\mu_i \right)^2}{2\sigma_i^2}}
\end{equation}
is a Gaussian with amplitude, mean, and standard deviation $A_i$, $\mu_i$ and $\sigma_i$, respectively. $i$ indicates the occupation of the dot, and $V$ is the measured voltage.

We define the measurement fidelity associated with occupation $i$ as
\begin{equation}\label{eq:fci}
f_{C,i} = \frac{\int_{V_s}^{V_f}g(V|A_i,\mu_i,\sigma_i)dV}{\int_{-\infty}^{\infty}g(V|A_i,\mu_i,\sigma_i)dV}
\end{equation}

\noindent
The integral bounds in Equation~\ref{eq:fci} are $V_s=-\infty$ and $V_f=V_t$ for $i=0$, and $V_s=V_t$ and $V_f=\infty$ for $i=1$, where $V_t$ is the threshold voltage.
$V_t$ is chosen to maximize the charge state readout fidelity
\begin{equation}\label{eq:Fc}
F_C = \frac{1}{2}\left(f_{C,0} + f_{C,1} \right).
\end{equation}

\noindent
Both $V_t$ and $F_C$ depend on the per-point integration time $T_{int}$. In our device, we achieve a charge state readout fidelity of $F_C=98.8\%$ and signal to noise ratio of $\sfrac{\left| \mu_1-\mu_2\right|}{\frac{\left( \sigma_1+\sigma_2 \right)}{2}}=4.3$ with an integration time as small as $T_{int}=100~$ns (Fig. 3d).
By extending the integration time to just $T_{int}=300$ ns, we achieve a charge state readout fidelity $F_C>99.9\%$ [Fig. 3c].

\begin{figure}[t!]
\includegraphics[width=8.6cm]{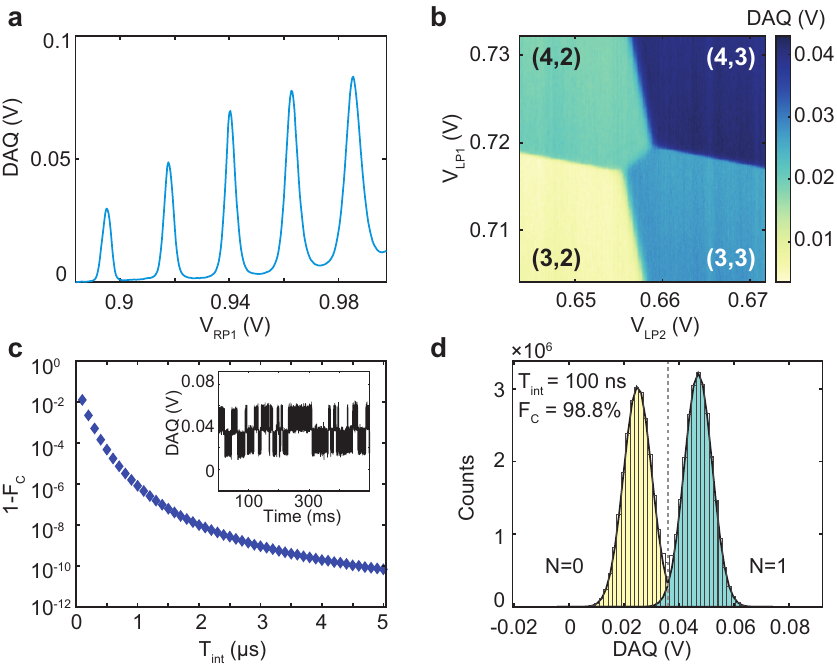}
\caption*{
\textbf{Fig. 3} Charge sensing via rf reflectometry.
\textbf{a}~Measurement of conductance peaks of the sensor dot using rf reflectometry.
%For charge sensing, $V_{RP1}$ is set on the side of a transport peak where the slope is largest.
\textbf{b}~Charge stability diagram of a double quantum dot acquired via rf reflectometry measurement of the sensor dot.
A plane has been subtracted to remove cross-talk between the double dot plunger gates and the sensor dot.
\textbf{c}~Plot of $1-F_C$ as a function of integration time $T_{int}$. 
The inset shows a representative time series with $T_{int}=300~\text{ns}$.
\textbf{d}~Histogram of the time series shown in \textbf{c} analyzed with an integration time $T_{int}=100~\text{ns}$ and having a measurement fidelity of $F_C=98.8\%$.
}\label{fig:chrg}
\end{figure}

\section{Singlet-Triplet Readout}

\begin{figure}[t!]
\includegraphics[width=8.6cm]{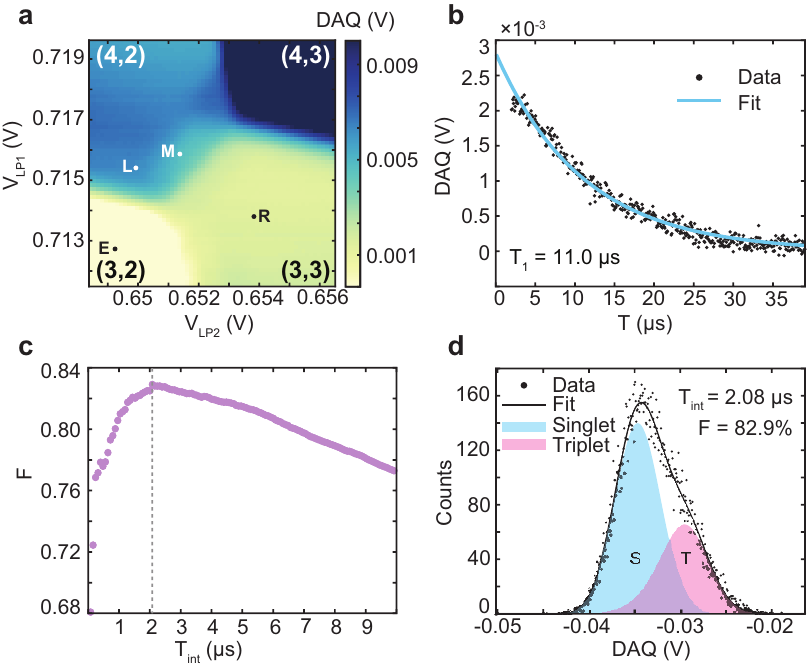}
\caption*{
\textbf{Fig. 4} Singlet-triplet readout via Pauli spin blockade.
\textbf{a}~Average charge sensor signal acquired at each measurement point across many repetitions of a pulse sequence that initializes a random spin state prior to pulsing to the measurement point.
A trapezoid near the interdot transition in the (4,2) charge configuration indicates the Pauli spin blockade region.
The limits of the color bar have been adjusted such that the signal in the (4,3) and (3,2) charge configurations is saturated so that the spin blockade region is more easily visible.
Positions L, R, E, and M, are the ground state initialization, random state initialization, empty, and measure positions, respectively.
\textbf{b}~Difference of the average signals measured at position M after initializing a random state and after initializing a singlet state as a function of measurement time exhibiting a decay with $T_1=11.0~\mu$s.
\textbf{c}~Plot of the singlet-triplet readout fidelity as a function of integration time.
A dashed line indicates the integration time at which the maximum fidelity is achieved.
\textbf{d}~Histogram of single shot measurements with an integration time of $T_{int}=2.08~\mu$s and a fidelity of $F = 82.9\%$.
The signal acquired on the DAQ is inverted from the signal shown in \textbf{a}, resulting in the singlet states having the lower voltage signal.
The data shown in \textbf{a}, \textbf{b}, \textbf{c}, and \textbf{d} were taken at $B_{ext}=50$~mT, where $B_{ext}$ is the externally-applied in-plane magnetic field.
}\label{fig:SB}
\end{figure}

Having demonstrated fast high-fidelity charge sensing, we turn to fast readout of spin states.
% capability we demonstrate here enables the possibility of fast single-shot spin state readout.
%In the remainder of this work, we turn our attention to reading out single and two-electron spin states on microsecond time scales.
%We read out the singlet and triplet states of a double quantum dot mapping spin states to charge states via conventional Pauli spin blockade, as well as a previously demonstrated enhanced charge state latching mechnism ~\cite{harvey2018high, studenikin2012enhanced, mason2015metastable}.
%As spin selective tunneling is widely used to measure spins of single electrons in silicon, we also study the performance this readout method using rf reflectometry.
%Spin blockade between the singlet and triplet states of a double quantum dot~\cite{johnson2005singlet, borselli2011pauli} is often difficult to observe in silicon based quantum dots~\cite{jones2019spin} because the valley degree of freedom can result in a small energy splitting of the lowest laying energy states such that the two electron ground state is nearly degenerate~\cite{boykin2004valley, tagliaferri2018impact}.
%From magnetospectroscopy, we estimate the valley splitting in our device to be $E_{V} \leq 20~\mu$eV.
%To avoid the lifting of spin blockade due to this small valley splitting, we tune our device to the (4,2)-(3,3) transition, thus fully occupying the lower valley energy levels~\cite{west2019gate, higginbotham2014coherent}.
We observe Pauli spin blockade at the $(4,2)-(3,3)$ transition in this device. (We did not observe spin blockade at the $(1,1)-(0,2)$ transition, likely because of a small valley splitting.) We repeatedly apply a three-step pulse sequence that initializes a random spin state in (3,3) prior to pulsing toward the measurement point, at which point the rf excitation is applied to the sensor dot.
When the randomly loaded spin state is a singlet, it can tunnel freely from (3,3) to (4,2). If it is a triplet, it remains blockaded in (3,3) until it either undergoes a spin flip or exchanges electrons with the reservoirs. We vary the position of the measurement point, and plot the average signal acquired at each measurement point in Fig. 4a. A trapezoid indicating the spin blockade region is visible in the (4,2) charge configuration near the interdot transition. 

To quantify the singlet-triplet readout fidelity, we perform 10,000 single-shot measurements in which we initialize a random spin state before pulsing to the measurement point in the spin blockade region, and an additional 10,000 measurements in which we instead preferentially initialize a singlet state prior to measurement.
The two sequences described above pulse the gates between positions E, R, and M in Figure 4a, and positions E, R, L, and M, respectively.
At the measurement point, we acquire a time series of the reflected rf signal for $40~\mu$s for each single-shot measurement. The average difference between these signals as a function of measurement time from 0-40 $\mu$s is shown in Fig. 4b.  These data follow a characteristic exponential decay with a relaxation time $T_1=11.0~\mu$s.
In fitting this data, we discard the first 2.5~$\mu$s of data at the beginning of each measurement to allow for the circuit to ring up. The value of $T_1$ that we measure is lower than typical spin relaxation times in Si based quantum dots~\cite{zajac2016scalable}.
This fast relaxation is likely related to the rf excitation and strong coupling between the dots and reservoirs in this device.  

To compute a measurement fidelity, we use Equations (1)-(3) of Ref.~\cite{barthel2009rapid} to fit a histogram of the data from the first pulse sequence discussed above to the sum of two noise broadened peaks with additional terms to account for relaxation~\cite{barthel2009rapid}. 
We extract the readout fidelity as $F = \frac{1}{2}( F_S +F_T)$, where $F_S$ and $F_T$ are the singlet and triplet readout fidelities~\cite{barthel2009rapid}. We choose the singlet-triplet threshold voltage to maximize the overall fidelity. Despite the enhanced triplet to singlet relaxation, we achieve a maximum fidelity of $F=82.9\%$ in $T_{int}=2.08~\mu$s (Fig. 4d). In this approach, we discard the first 500 ns of data to allow the resonator to ring up. The total measurement time is thus $2.58~\mu$s. In computing the fidelity, we have accounted for spin relaxation during this 500-ns interval. 

To improve our readout fidelity, we use a charge latching mechanism~\cite{studenikin2012enhanced, mason2015metastable, harvey2018high}. We tune our device such that the tunneling rate between the dot under LP1 and its corresponding reservoir is $\Gamma_1\sim10~\text{MHz}$ and the tunneling rate between the dot under LP2 and its corresponding reservoir is $\Gamma_2\ll\Gamma_1$.
%the lead connecting the dot under LP1 to its reservoir is very fast, while the lead connecting the dot under LP2 to its reservoir is very slow.
In this tuning, we again apply a three-step pulse sequence that loads a random state prior to pulsing to the measurement point, and vary the measurement point (Fig. 5a).
The charge latching mechanism allows singlet states to tunnel across to (4,2), but triplet states instead preferentially tunnel to an excited charge state in (4,3). 
%maps singlet and triplet states to (4,2) and (4,3) charge occupancies, respectively.
Generally, this technique results in better sensitivity than conventional spin-blockade readout, because the total electron number differs between these states. 

We characterize the readout by performing 10,000 single-shot measurements at position M in Fig.~5a after initializing a random state by pulsing to positions E and then R, as well as after initializing a singlet state by pulsing to positions E, R, and then L.
We observe a longer decay time of $T_1 = 51.9~\mu$s [Fig. 5b], which is likely due to the reduced electron exchange rate with the reservoir connected to the double dot via the slow tunnel barrier.
We compute the fidelity as before, but we now subtract an additional mapping error~\cite{harvey2018high} $e_{map}=\frac{1}{T_{int}}\int_{T_0}^{T_0+T_{int}}e^{-\sfrac{t}{T_{L}}}dt$. Here, $T_L \approx 150$ ns is the average tunneling time across the tunnel barrier connecting the dot under LP1 to its reservoir, and $T_0=650 $ ns is the time we discard once the rf excitation is applied to allow the resonator to ring-up and the latching process to take place.  $e_{map}$ is the average probability during the integration time that a triplet will not have tunneled to the $(4,3)$ state and will be mistakenly identified as a singlet. Figure 5c shows the fidelity as a function of integration time for this method. We achieve a fidelity $F>98\%$ with an integration time as short as $T_{int}=800~\text{ns}$ [Fig 5d], and an improved maximum fidelity of $F=99.0\%$ in $T_{int}=1.65~\mu$s [Fig 5c].

\begin{figure}[t!]
\includegraphics[width=8.6cm]{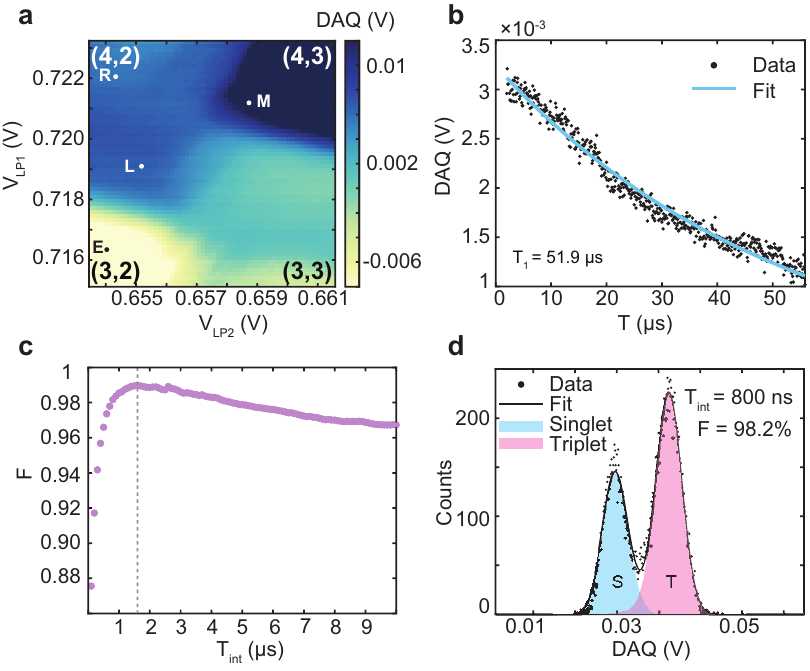}
\caption*{
\textbf{Fig. 5} Singlet-triplet readout utilizing a latching mechanism.
\textbf{a}~Average charge sensor signal acquired at each measurement point across many repetitions of a pulse sequence that initializes a random spin state before pulsing to the measurement point.
Positions L, R, E, and M, are the ground state initialization, random state initialization, empty, and measure positions, respectively.
\textbf{b}~Difference of the average signals measured at position M after initializing a random state and after initializing a singlet state as a function of measurement time exhibiting a decay with $T_1=51.9~\mu$s.
\textbf{c}~Singlet-triplet readout fidelity as a function of integration time.
A dashed line indicates the integration time at which the maximum fidelity is achieved.
\textbf{d}~Histogram of single shot measurements with an integration time of $T_{int}=800~$ns and a fidelity of $F = 98.2\%$.
The data shown in \textbf{a}, \textbf{b}, \textbf{c}, and \textbf{d} were taken at $B_{ext}=50$~mT.
}\label{fig:latched}
\end{figure}

\section{Single Spin Readout}

\begin{figure}[t!]
\includegraphics[width=8.6cm]{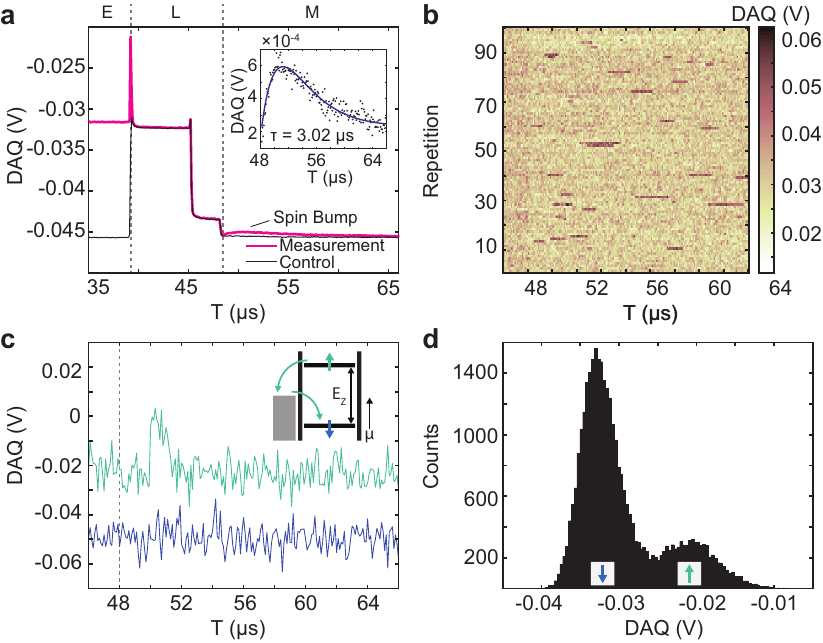}
\caption*{
\textbf{Fig. 6} Single spin readout.
\textbf{a}~Averaged DAQ signal during the single spin readout pulse sequence (pink). 
The sequence empties the dot, loads an electron with a random spin, and then pulses to the measurement point.
Empty, load, and measurement sections of the pulse are separated by dashed gray lines, and are labeled with E, L, and M, respectively, above the plot.
A spin bump corresponding to spin up tunneling events is visible.
A control pulse that initializes a spin down electron prior to measurement is shown in black.
%The dashed gray line at $T=48~\mu$s indicate when the single-shot measurement time begins.
The difference between the control and measurement pulses during the measurement window is shown in the inset.
\textbf{b}~100 single-shot spin selective tunneling measurements demonstrating individual electron tunneling events.
The plunger gate is pulsed to the measurement point at $T=48~\mu$s in the pulse sequence.
\textbf{c}~Two representative single shot traces. The top trace shows a spin-up electron tunneling out and back in as a spin down. The bottom trace shows no spike and indicates a spin-down electron. The traces have been offset for clarity.
\textbf{d}~Histogram of the maximum single-point value of the signal in the range $T=48~\mu$s to $T=66~\mu$s.
The data shown in \textbf{a}, \textbf{b}, \textbf{c}, and \textbf{d} were taken at $B_{ext}=1.5$~T and a sampling rate of 10~MHz.
}\label{fig:sst}
\end{figure}

Having demonstrated fast, high-fidelity singlet-triplet state readout, we now discuss fast single-spin readout via spin-selective tunneling~\cite{elzerman2004single}.
Operating the device near the (0,0)-(1,0) transition and with $B_{ext}=1.5$~T, we apply a three step pulse sequence~\cite{elzerman2004single} to plunger gate LP1 which empties and then loads the corresponding dot with a random spin, and then pulses to the measurement point (Fig. 6a). At the measurement point, which is close to the (1,0)-(0,0) transition, a spin-up electron will preferentially tunnel out. Some time later, a spin-down electron will tunnel back in. This brief change in occupancy results in a measurable change in the charge-sensor signal.

We acquire more than 32,000 single-shot measurements using the pulse sequence described above.
For each single-shot measurement, we additionally perform a control measurement using a pulse sequence in which we initialize a spin-down electron instead of an electron with a random spin state.
Figure 6a shows plots of the average acquired charge sensor signal across all single-shot traces for both the measurement and control pulse sequences.  
A ``spin bump" from the presence of tunneling events corresponding to spin up electrons is visible at the beginning of the measurement window ranging from $T=48-66~\mu$s. The signal from the control pulse shows no spin-bump, as expected.
The inset of Fig. 6a shows the difference between the average of the control and measurement pulse sequences in the measurement window.
We fit these data to a function of the form $b(T)=A+B(T/\tau)e^{-T/\tau}$, where $A$, $B$, and $\tau$ are a fit parameters, and $T$ is time from the start of the measurement window. We extract a characteristic tunneling time $\tau=3.02~\mu$s. A representative collection of 100 single-shot traces is shown in Figure 6b, and two traces (one spin up and the other spin down) are shown in Fig. 6c.

%Individual representative traces corresponding to spin up and spin down electrons are shown in Figure 5c.
Figure 6d shows a histogram of the maximum single point value acquired in each single-shot measurement during the measurement window. This histogram shows two distinct peaks corresponding to spin-up and spin-down electrons. The overall acquisition time for each single-shot measurement is only 18 $\mu$s, orders of magnitude faster than usual spin-selective tunneling measurement times~\cite{elzerman2004single}. This increase in speed is enabled by the high bandwidth of the reflectometry circuit.

\section{Conclusion}

We have optimized a Si/SiGe quantum dot device with overlapping gates for rf reflectometry by making only modest geometric changes to our device design.
The methods we use are applicable to Si devices with and without overlapping gates, can be implemented with relative ease, preserve the scalability of the gate layout, and, importantly, provide the ability to perform rapid high-fidelity charge and spin state readout.
We have demonstrated microsecond-scale readout of single-spin states and sub-microsecond singlet-triplet readout. We expect that further improvements are possible via optimization of the dot-reservoir couplings and the sensor dot position. This work presents a feasible solution to achieving rapid and high-fidelity spin-state readout in Si spin qubits that is largely compatible with existing device designs.

During the completion of this manuscript, we became aware of a related result demonstrating similar techniques for the implementation of rf reflectometry in accumulation mode Si devices~\cite{noiri2020radio}.

We thank Aaron Mitchell Jones for valuable discussions.
We thank Lisa F. Edge of HRL Laboratories, LLC. for the epitaxial growth of the SiGe material.
Research was sponsored by the Army Research Office and was accomplished under Grant Numbers W911NF-16-1-0260 and W911NF-19-1-0167.  The views and conclusions contained in this document are those of the authors and should not be interpreted as representing the official policies, either expressed or implied, of the Army Research Office or the U.S. Government.  The U.S. Government is authorized to reproduce and distribute reprints for Government purposes notwithstanding any copyright notation herein.
E.J.C. was supported by ARO and LPS through the QuaCGR Fellowship Program.

%\bibliography{support/bib}

\end{document}